\documentclass[graybox]{svmult}

\usepackage{type1cm}        
\usepackage{makeidx}         
\usepackage{graphicx}        
\usepackage{multicol}        
\usepackage[bottom]{footmisc}

\usepackage{newtxtext}       %
\usepackage[varvw]{newtxmath}       

\makeindex             

\usepackage[numbers,sort]{natbib}
\makeatletter
\renewcommand\@biblabel[1]{#1.}
\makeatother

\begin{document}

\title*{Towards Data-Driven Modeling of Cell Cycle and Wound Closure Processes}

\titlerunning{Data-Driven Modeling of Cell Cycle and Wound Closure}

\author{Erik~Blom\orcidID{0009-0005-8141-6802}, Qiyao~Peng\orcidID{0000-0002-7077-0727}, Leah~Pomfret\orcidID{0009-0005-1925-3299}, Richard~Mort\orcidID{0000-0002-3566-8217} and Stefan~Engblom\orcidID{0000-0002-3614-1732} }

\institute{Erik Blom \at Division of Scientific Computing, Department of Information Technology, Uppsala University, Uppsala, Sweden. \email{erik.blom@it.uu.se}
\and Qiyao Peng (Correspondence) \at (1) Mathematics for AI in Real-life Systems, Schools of Mathematical Sciences, Lancaster University, Lancaster, United Kingdom; (2) Mathematical Institute, Leiden University, Leiden, The Netherlands. \email{qiyao.peng@lancaster.ac.uk}
\and Leah Pomfret, Richard Mort \at Division of Biomedical and Life Sciences, Lancaster University, Lancaster, United Kingdom. \email{l.pomfret@lancaster.ac.uk, r.mort@lancaster.ac.uk}
\and Stefan Engblom \at (1) Division of Scientific Computing, (2) Science for Life Laboratory, Department of Information Technology, Uppsala University, Uppsala, Sweden. \email{stefan.engblom@it.uu.se}
}

%
%
\maketitle


\abstract{
  Effective wound repair treatments rely on a clear picture of how cell proliferation and migration are coordinated during tissue restoration. Fibroblasts are key contributors to tissue restoration in the dermis, and modern imaging tools allow their cell-cycle progression to be observed directly, enabling comparison between experiments and computational models. Here we investigate how different stages of the cell cycle influence fibroblast-driven wound closure using the Discrete Laplacian Cell Mechanics (DLCM) framework driven by time-lapse microscopy data. \textit{In vitro} assays provide cell positions, migration behaviour, and cycle-stage information, and we show that incorporating proliferation, migration, and cell cycle arrest allows the computational model to reproduce the essential experimental trends. The results reveal that arrest in the G1 phase notably impacts the cell cycle dynamics and that the initial spatial arrangement of cycle states significantly affects wound closure. By linking single-cell cycle dynamics with emergent tissue behaviour this work establishes a quantitative approach for exploring how intracellular processes shape repair processes. More broadly, it demonstrates the value of integrating high-resolution data with cell-based mechanical models and provides a foundation for systematic \textit{in silico} evaluation of therapeutic interventions.
}

\section{Introduction}
\label{sec:intro}

The cell cycle is fundamental to the development and maintenance of multicellular organisms --- driving embryonic growth, tissue renewal and wound healing \cite{Vermeulen2003}. Generally speaking, there are four ordered phases in a complete cell cycle, namely G1, S, G2 and M. G1 represents the first growth phase, during which the cell grows in size and prepares for DNA replication, e.g. mRNA and protein synthesis. This is generally the longest phase of the cell cycle, comprising $30--40\%$ of its total time. Before a cell enters S phase and replicates its DNA, it must traverse the first internal checkpoint (the so-called G1 checkpoint or restriction point). After S phase, the cell enters a further growth phase --- G2. The G2 checkpoint ensures DNA integrity and the cell may pause here to repair any erroneous DNA; otherwise, the cell first segregates its chromosomes through a process called mitosis, followed by cytokinesis to generate two identical daughter cells. For more details concerning the cell cycle, we refer to \cite{Alberts2002MolecularBiology, Schafer1998, Vermeulen2003}.


Wound healing is the process by which the skin heals itself after injuries, and is generally divided into four stages. In this paper we are mostly concerned with the third stage, known as the \emph{proliferation phase}, where fibroblasts proliferate and collectively migrate towards the wound to reconstruct the damaged tissue \citep{Addis2020, Mamun2024, Cen2022}. The coordination between the proliferation and migration of the fibroblasts plays a crucial role in the quality of wound healing, a coordination which naturally involves cell cycle activity \citep{VandeBerg2003}. Improper cell cycle functioning of fibroblasts can cause pathological healing such as chronic wounds \citep{VandeBerg1998}. Therefore, it is of great interest to better understand the correlations between the cell cycle of fibroblasts and wound healing, particularly for the development of new treatments.

Cell cycle biosensors allow the monitoring of cell proliferation in live cells using time-lapse microscopy. The best known is the Fucci(SA) probe pair, composed of truncated forms of two cell cycle-regulated proteins --- CDT1 and Geminin (GMNN) --- fused to red and green fluorescent proteins, respectively. CDT1 is a DNA replication licensing factor that accumulates in G1 and is targeted for degradation in S/G2/M phase, ensuring DNA replication occurs only once per cell cycle. GMNN does the opposite: it is degraded in G1 but accumulates in S/G2/M phase. As a result, Fucci(SA) labels G1 cells red and S/G2/M cells green, allowing direct visualisation of cell cycle progression in living tissue.

Fucci2a is a genetically encoded, dual-colour fluorescent cell‑cycle reporter that utilises a self-cleaving peptide to express the components of the Fucci(SA)2 \cite{Abe2013} probeset at equimolar concentrations \cite{Fucci2a}. In practice, this enables tracking cell-cycle progression and proliferation dynamics simply by imaging red and green nuclear fluorescence. This opens up the possibility for recording highly detailed data streams of wound healing processes, thus opening up for building data-driven computational models based on first principles. The main targets with this endeavour are to reveal the underlying biomechanics, test hypothesis and predict the consequences of various interventions \cite{brodland2015computational}. Since population heterogeneity and cell-specific detail need to be resolved here, \emph{cell-based models} are a natural choice. Such frameworks can treat each cell (or possibly other biological units) as an individual with distinct properties, enabling, e.g., the tracking of precise location and the various cellular activities. A variety of cell-based models have been developed in the past, describing different layers of the skin and stages of wound healing \citep{Peng2020, Zhao2017, Boon2016, Wang2019}.

In this paper, the aim is to build a computational model which facilitates an investigation of the correlation between the cell cycle and the wound closure process. To this end, we will be relying on the Discrete Laplacian Cell Mechanics (DLCM) framework, introduced in \cite{Engblom_2018}, and developed further in \cite{blom2025dlcm}. To strengthen the study, we compare simulation results from the DLCM model with \textit{in vitro} experiments of NIH 3T3 mouse embryonic fibroblasts labelled with Fucc2a, generating data on individual cell cycle stages alongside cell position and migration. These experiments use a polydimethylsiloxane (PDMS) insert to create a well-defined wound in a confluent cell monolayer, avoiding the cell damage associated with traditional scratch assays to minimise confounding signalling processes, facilitating a closer comparison between model and experiment.

The manuscript is structured as follows: Sect.~\ref{sec:experiment} describes how the laboratory experiments are set up and the resulting data. The model, extended from the DLCM framework, is explained in Sect.~\ref{sec:model} and simulation results are shown in Sect.~\ref{sec:results}. Finally, conclusions and discussion are delivered in Sect.~\ref{sec:conclusions}.

\section{Experiment Setup}\label{sec:experiment}
NIH 3T3 mouse embryonic fibroblasts expressing the Fucci2a biosensor were employed to visualise cell cycle progression during live imaging. Cells were maintained in Dulbecco’s Modified Eagle’s Medium supplemented with $10\%$ fetal bovine serum, $1\%$ penicillin-streptomycin, and $100 \mu g/mL$ hygromycin. Cells were incubated at $37^{\circ}C$ at $5\%$ CO$_2$ in a humidified incubator and were passaged when $70-80\%$ confluent.

To simulate wounds, PDMS culture inserts were fabricated by cutting a PDMS sheet into $2mm\times1mm$ rectangular strips using a scalpel. Prior to cell seeding, a single insert was pressed into the centre of a glass-bottomed 24-well plate in each well to create a defined void region upon removal. Fucci2a NIH 3T3 fibroblasts were seeded at a density of $6 \times 104$ cells$/mL$ and incubated for 24 hours before removing the PDMS insert with fine-tipped tweezers. Cell cycle and cell migration were observed using a Zeiss LSM880 confocal microscope equipped with an incubation chamber maintained at $37^{\circ}C$ and $5\%$ CO$_2$. A definite focus module was used to maintain focal stability during long-term imaging and Z-stack images were acquired using a $20\times$ lens.

\section{Modeling Framework}
\label{sec:model}


The DLCM framework consists of two layers of description, namely, the cellular and the population level. The cellular level describes the intracellular dynamics that decide the cell fate or \emph{cell state}, e.g., whether a cell will proliferate. The population level handles the "consequence", e.g., one cell divides into two, including cell migration and the micro-environment that may affect the cell state, such as cellular pressure.

Denote $w_i$ as the cell state for cell $i$. The cell state is updated by applying a continuous-time Markov chain, with event propensity $\nu_r$ and stoichiometric matrix $\mathbb{S}$:
\begin{equation}
   \label{eq:Markov_chain_example_cell_state}
   w_i \xrightarrow[]{\nu_r} w_i + \mathbb{S}_r.
\end{equation}
Note that $\nu_r$ captures a combination of internal and external environmental signals.

At the population level, the DLCM framework represents cells as distinct fluid-like volumes that reside in a fixed lattice \cite{blom2025dlcm}. The spatial domain is discretized into $i = 1, 2, ..., N_{\mathrm{vox}}$ voxels. Each voxel may contain a number of cells, denoted as $u_i = 0, 1, 2$. The population is embedded in a micro-environment field, $v_i$, governed by quasi-stationary partial differential equations (PDEs) that may affect the cell dynamics. Cells may die, proliferate, or switch phenotype; the number of cells of a certain type in voxel $i$ changes according to the event
\begin{equation}
   \label{eq:Markov_chain_example_cell_number}
   u_i \xrightarrow[]{\omega_r} u_i + \mathbb{N}_r,
\end{equation}
where $\omega_r$ is the event propensity dependent on the cell state, the crowdedness of the surrounding voxels and the micro-environment.

We introduce two cell states for cell $i$, corresponding to the G1 phase, denoted by $g_i$ in the model, and the rest, i.e., S/G2/M, denoted by $m_i$. For brevity, we omit the index $i$ of these phases. Cells with the $g$ state proceed to $m$, and cells in the $m$ state divide into two cells with the $g$ state, given that certain conditions hold which we describee below. Here, \eqref{eq:Markov_chain_example_cell_number}, in voxel $i$, can be specified as 
\begin{equation}
    \label{eq:cycle_stages}
    \left\{
    \begin{aligned}
  g &\xrightarrow[]{\lambda (\tau_g=\kappa_g)\times(u_i<2)} m, \\
  m &\xrightarrow[]{\lambda (\tau_g=\kappa_g)\times(u_i<2)} g + g.
    \end{aligned}
    \right.
\end{equation}
Note that both cell state changing can only occur when the cell solely occupies voxel $i$. Here $\lambda$ is the rate of the corresponding Erlang distribution, and $\tau_{\{g,m\}}$ denotes the state of the sub-stage (or \emph{tick}), of the cell in the $\{g,m\}$-phases, respectively, which are incremented by the internal events
\begin{equation}
   \label{eq:cycle_ticker}
    \begin{aligned}
  \tau &\xrightarrow[]{\lambda ( \tau\leq\kappa)} \tau+1, \\
    \end{aligned}
\end{equation}
where we have omitted the subscripts $m$, $g$. We consider three different variants of additional boolean conditions on \eqref{eq:cycle_ticker} in \S\ref{sec:results_cellcycle}.
Note the ticker system represents an abstract progression through the cell cycle where each sub-stage $\tau$ does not necessarily correspond to a definite physical state. A similar cell cycle scheme is studied in \cite{Yates2017}, where they find that $\lambda = 0.0083$ and $k=12$, yield a cell cycle time of $k/\lambda=1445\min$ (i.e. roughly $24.09$ hours). Following this and assuming that the G1-phase takes up about $40\%$ of the full cycle, we define $\kappa_g = 4$ and $\kappa_m = 6$. The model parameters are summarized in Table~\ref{tab:parameter_value}.

As the cell fate (cellular level) has an impact on the cell behaviour (population level), the state change of the cells is coupled with migration, which is a pressure-driven Markov process between voxels $i$ and $j$:  
\begin{equation}
\label{eq:cell_migration}
  u_i \xrightarrow[]{(D\nabla p)_{ij}} u_j,
\end{equation}
where $p$ is the pressure field in the computational domain, derived from continuous physics according to Darcy's law for flow through porous media (i.e., the extracellular matrix (ECM)). The constant $D$ is flexible enough to scale the movement rates according to the number of cells in both voxel $v_i$ and $v_j$ as specified in Table~\ref{tab:parameter_value}. Darcy's law states that cell migration velocity is given by
\begin{equation}
  \label{eq:darcys_law}
  \boldsymbol{v} = -D \nabla p,
\end{equation}
with Darcy constant $D$ to be interpreted as the ratio of
the medium permeability $\kappa$ to its dynamic viscosity $\mu$, $D := \kappa/\mu$. The pressure $p$ is modelled by a stationary Laplace equilibrium:
\begin{equation}
\left\{
 \begin{aligned}
  \label{eq:pressure_law}
 -\Delta p &= s(u) + r(u), \quad &\mathrm{in} \; \Omega(t),\\
 p &= 0, \quad &\mathrm{on} \; \partial\Omega^1(t),\\
 \end{aligned}
 \right.
\end{equation}
with $s(u) = 1$ when the voxel contains two cells in the $g$ state and zero otherwise, and $r(u)$ is equal to the number of cells in the $m$ stage in that voxel. As a consequence, cells primed to divide always exert cell pressure to prepare space for the daughter cells. $\Omega(t)$ is the union of voxels which contain cells.

\begin{figure}[h!]
\centering
  \includegraphics[width=1\linewidth]{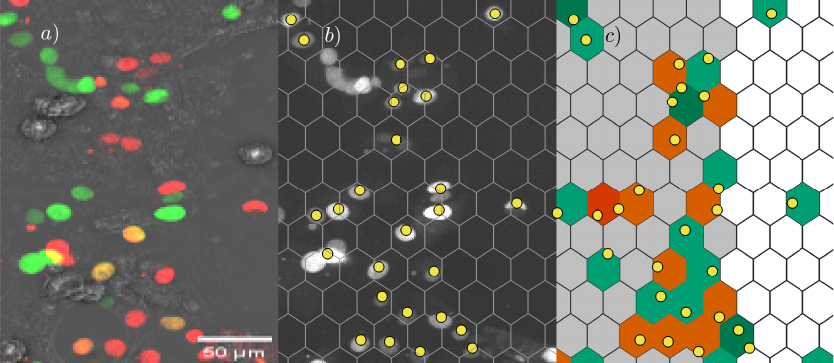}
  \caption{From left to right: \textit{a)} cell data with red representing the G1 phase, green--S/G2/M; \textit{b)} cell data in black and white and overlayed with a hexagonal mesh and cell center (yellow points) where the image processing finds a cell; \textit{c)} finally, the DLCM model's representation formulated from the number of cells in each voxel given by the data. Green voxels contain cells in $g$ state; red voxels contain cells in $m$ state (cell states randomized as described in \S\ref{sec:results_setup}). Darker-shaded voxels indicate that the voxel is doubly occupied (states of the two cells may differ but we do not specifically use another color for the visualization). The white and grey voxels, where there are no cells, represent wounded and unwounded region, $W$ and $V$, respectively.}
 \label{fig:data2dlcm}
\end{figure}

For the boundary conditions we split the domain of empty voxels into two non-overlapping domains: wound and non-wound, denoted by $W(t)$ and $V(t)$, respectively, the former comprising all voxels that have never been occupied. Specifically, $V(t)$ is the union of the domains $\Omega(\tau)$ for $\tau \in [0,t]$ and a certain initial domain $V(0)$, essentially updating the non-wound domain with the information about where cells \emph{have been}. We initialize $V(0)$ as the domain left of the initial population boundary such that the wound is to the right, with the boundary defined at the mean position in $x$ of the foremost ten cells --- see Fig.~\ref{fig:data2dlcm}\emph{c)} where the grey voxels are to the left of the wound boundary. For the boundary $\partial\Omega^1(t)$ against the wound $W$ we apply a homogeneous Dirichlet boundary condition for the pressure, whereas for the boundary against $V$, a natural Neumann condition for the pressure is applied. This modelling choice stems from the fact that fibroblasts release ECM components and the cell cycle process requires the availability of ECM \cite{VandeBerg2003}, which is thus only available in unwounded region or part of the wounded area where cells have visited.

\section{Numerical Results}
\label{sec:results}

In this section, we begin by describing how the model initial conditions are informed by the data as well as how certain parameters are calibrated in Sect.~\ref{sec:results_setup}. The main emphasis lies in comparing the dynamics of the model cell cycle with the data and how the model responds to cell cycle arrest which we do in Sect.~\ref{sec:results_cellcycle}. Finally, we consider different initial spatial patterns of the cell cycle phases and their impacts on the wound closure rate in Sect.~\ref{sec:results_cycle_pattern}.


\subsection{Simulation Setup}
\label{sec:results_setup}

From the data we have access to the number of cells with the $g$ and $m$ state and cell positions. Regarding the initial configuration of the cells, we map the position of the cells in the experimental data to an $N\times N$ hexagonal mesh by counting the number of cells contained inside each voxel; see Fig.~\ref{fig:data2dlcm}. We choose $N$ such that maximum two cells are allowed in each voxel. All data samples have been preprocessed in the same way and during the experiments we uniformly sample one of them as an initial condition. We use the number of cells in $g$ and $m$ state from experimental data to inform the model states, and uniformly sample that number of cell indices to be assigned $g$ and $m$, respectively. This means that we are investigating the assumption that there is no distinct pattern in the initial spatial configuration of cell states in the biological experiments. In \S\ref{sec:results_cycle_pattern}, we study the impact of two distinct initial spatial patterns of cell states on the model wound closure rate. The initial tick is also uniformly sampled.

\begin{figure}
\includegraphics[width=1\linewidth]{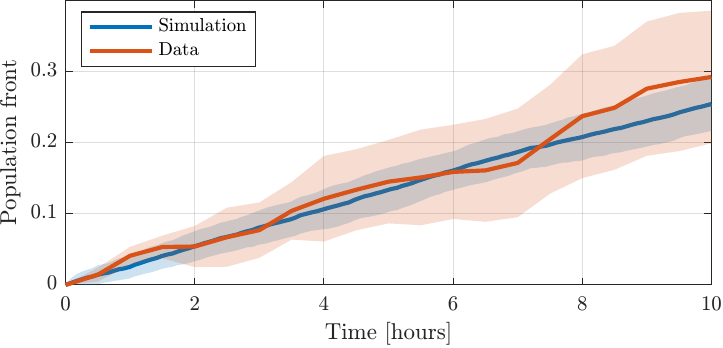}
\caption{Estimated front position of the population along with sample standard deviation of $n = 100$ simulations. Fitting a line through either the data or the simulation results yields a slope of $0.03$, which is also how the Darcy scaling coefficient $D$ in \eqref{eq:cell_migration} was selected.}
 \label{fig:population_front}
\end{figure}

The parameter values are found in Table~\ref{tab:parameter_value}. While using the parameter values from the experiments and prior knowledge as much as possible, some parameter values can only be estimated. The travelling speed of the wound front is used to estimate the parameter $D$ in \eqref{eq:cell_migration} which represents the mobility of the cells. With the set of parameter values in Table~\ref{tab:parameter_value}, the wound front comparison is shown in Fig.~\ref{fig:population_front}.
\begin{table}[h!]\footnotesize
    \centering
    \caption{Parameter values of the model.}
    \begin{tabular}{c|c|c|l}
    \hline
      \textbf{Parameter}  &  \textbf{Value} & \textbf{Dimension} &\textbf{Description}\\
      \hline
      $\lambda$ & $0.0083$ & $[t^{-1}]$ & Rate of increase in the cell sub-stages $\tau_{\{g,m\}}$. \\
       $\kappa_g$ & $4$ & - & Threshold of $\tau_g$ for progression from phase $g$ to $m$.\\
       $\kappa_m$ & $6$ & - & Threshold of $\tau_m$ for progression from phase $m$ to $g$. \\
       $D$ & $1$ & $[(tf)^{-1}]$ & Migration rate scaling into non-crowded voxels. \\
       \hline
    \end{tabular}
    \label{tab:parameter_value}
\end{table}

The baseline model describes the healing process of a wound (as in Sect. ~\ref{sec:experiment}), where the wound is located at the right side of the computational domain. To mimic the function of $G1$ checkpoint in the cell cycle, we only allow the cell cycle arrest in $g$ state but not in $m$ state. 

\subsection{Cell cycle}
\label{sec:results_cellcycle}


We firstly investigate the emergent dynamics of the cell cycle as it couples to the population-level behaviour, in part through the blocking of the cycle progression.
We consider three cases regarding the cell cycle modelling, specifically variants of arresting the ticker progression in \eqref{eq:cycle_ticker}:
(i) the cell cycle always proceeds further; (ii) both G1 and G2 checkpoints are involved, i.e. cell cycle progression pauses when they are in a doubly occupied voxel (i.e., adding the boolean $u_i<2$ to the event rate for both cell states); (iii) only G1 checkpoint is modelled, i.e. only cells in $g$ state halt the cell cycle progression when they are in a doubly occupied voxel, whereas the cell cycle always proceeds when they are at $m$ state. The case (iii) is the default setting in our simulations. Recall that cell state changes ($g$ to $m$ or cell division) still only occur when the cell solely occupies the voxel. 

\begin{figure}[h!]
\includegraphics[width=1\linewidth]{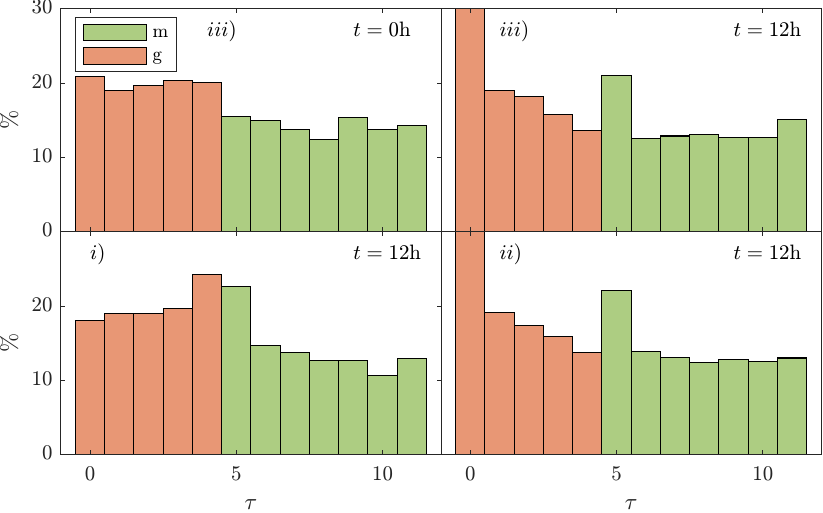}
\caption{Percentage of cells from a total of $n=100$ simulations at each sub-stage (ticker) for respective cell cycle stage ($g$ and $m$) at the start and end of the simulation for case (iii), where $g$ cells in doubly occupied voxels do not tick. We also show the end of the simulation for case (i) and (ii). Here, green and red bars represent $m$ and $g$ state, respectively.}
\label{fig:tickers}
\end{figure}

We present the distribution of the cell cycle sub-stages at the end of the simulation for the three cases in Fig.~\ref{fig:tickers}(i)--(iii), respectively (with the initial uniform distribution in Fig.~\ref{fig:tickers} upper left panel). In general, when the cell cycle is arrested (case (ii) and (iii)), most cells are in the early stage of $g$ (in line with the theoretical results in \cite{Yates2017}), a phase-alignment emerging since one $m$ cell divides into two $g$ cells; It is possible that this effect is exaggerated in our model since right after the division, two $g-$state cells are in the same voxel, which delays further progression. In case (i), the cells tend to aggregate in the latter stage of $g$ state, since the state change is only allowed when the voxel is solely occupied.

In this study, the impact of cell-cell interaction on the cell cycle is taken into account. We found no significant difference between case (ii) and (iii), in which the latter allows $g$-cells to progress freely.  Hence, the cell cycle arrest at the $g$ state has a significant influence on the distribution of the cell cycle stages, clearly discernible from arrest during the $m$ state.

\begin{figure}[h!]
\centering
\includegraphics[width=1\linewidth]{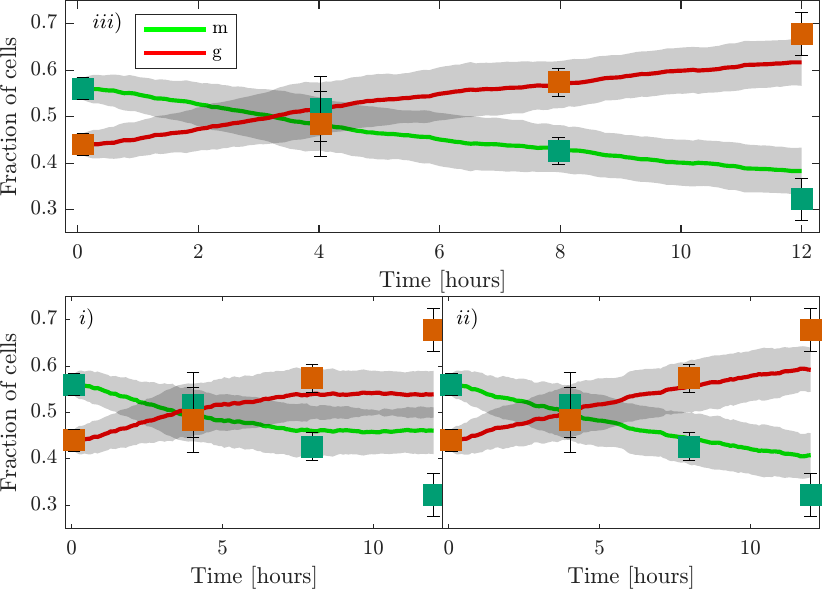}
 \caption{Fraction of cells in respective cycle stage, showing all three cases for the cell cycle arrest. The panels represent cases (i)-(iii), respectively. Red and green are the $g$ and $m$ states (or G1 and S/G2/M), respectively; the box plots are the experimental data, and the curves and envelopes represent the simulation mean and 68\% confidence interval of $n = 100$ simulations.}
 \label{fig:cycle_iii}
\end{figure}

\noindent
To further understand the composition of cells at various stages in the cell cycle and better use the limited experimental data, instead of the number of cells, we focus on the proportion of cells at four time points, as is shown in Fig.~\ref{fig:cycle_iii}. Again, we vary the cell cycle arrest cases as mentioned previously. For case (i) where there is no cell cycle arresting (Fig.~~\ref{fig:cycle_iii}\textit{a)}), the proportion of cells at different states tends to become steady after $6$ hours and the difference between the portions of $g$ and $m$ states is significantly smaller than the experimental data. With the cell cycle arrest included in the model (case (ii) and (iii) in Fig.~\ref{fig:cycle_iii}\textit{b)} and \textit{c)}, respectively), the simulation results agree with the experimental data well, which highlights the importance of modelling the cell cycle arrest. However, there remains a small discrepancy at the final time step, which might be due to regular exposure to the flash of the camera, causing irregular behaviour towards the end.
 
\begin{figure}[h!]
\centering
\includegraphics[width=1\linewidth]{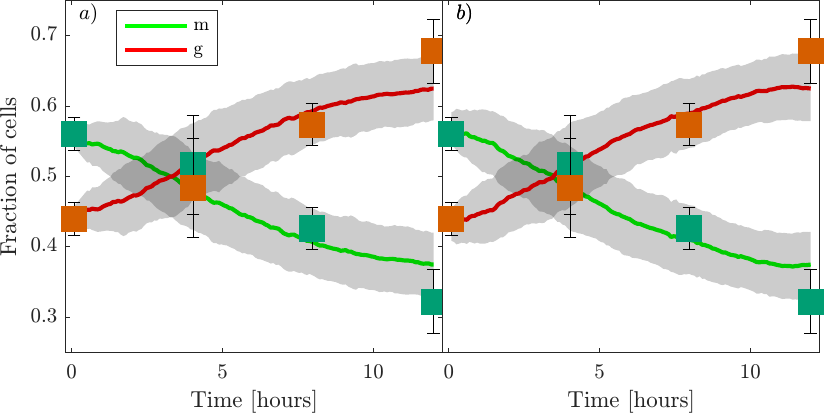}
 \caption{Fraction of cells in respective cycle stage, showing both simulation and data mean and standard deviation of $n = 100$ simulations, with colour scheme as in \ref{fig:cycle_iii}. The model is case iii) but using different spatial distribution of the initial cell cycle phases: \emph{a)} The leftmost cells are in $g$; \emph{b)} The rightmost cells are in $g$. The most notable difference in behaviour occurs initially.}
 \label{fig:cycle_spatial}
\end{figure}

\subsection{Initial spatial distribution of cell state}
\label{sec:results_cycle_pattern}
\noindent
Due to a lack of information from the experimental data regarding the spatial pattern of cell states, we randomly sampled the state and the initial ticker in the above experiments. To investigate the impact of the cell state spatial distribution on wound closure, we consider two spatial pattern extrema: (a) cells closest to the wound are in the $g$ state and the rest in the $m$ state are further away; and (b) vice versa.
The cells in the $m$ state are nearer to proliferation which, as they are also closer to the wound, benefit from the increased availability of space there and we might expect an increase in the closure rate. In other words, the acceleration of wound closure is a direct combination of both cell migration and proliferation. However, in case (b), when cells in the $m$ state are further away from the wound edge, the proliferation of cells tends to be delayed until the front cells in $g$ state have migrated towards the wound to make space. Hence, for case (b), wound closure initially relies on the cell migration other than proliferation.

We apply Wilcoxon signed-rank test on simulation results to determine whether the wound closure rates vary as a consequence of the spatial distribution of the cell state, i.e., cases (a) and (b). We measure wound closure by the position of the front, which is moving towards the right. Denote $r$ as the average speed of the wound front, hence, a larger $r$ indicates a faster wound closure. The statistical test results are shown in Table~\ref{tab:wilcox_test_closure_rate}. All the $p-$values are very small, which suggests a significant difference between the wound closure rate due to the initial spatial distribution of the cell states. For reference we state the mean and standard deviation of the final front position for the three cases in order of ascending magnitude: case (a): $0.207 \pm 0.041$; baseline (cf.~Fig.~\ref{fig:population_front}): $0.256 \pm 0.038$; case (b): $0.291 \pm 0.033$. Further investigation from both laboratory and modelling perspectives is needed for future study.
\begin{table}[h!]
    \centering
    \caption{Wilcoxon signed-rank test results for comparisons of the average speed of the wound front $r$ (which reflects the wound closure rate) under different initial conditions. All alternative hypotheses are one-sided.} 
    \label{tab:wilcox_test_closure_rate}
    \begin{tabular}{l|r|r|r}
      \hline
      Alternative hypothesis & $r({\rm m\ right})>r({\rm random})$ & $r({\rm random})>r(\mathrm{g\ right})$ & $r(\mathrm{m\ right})>r(\mathrm{g\ right})$\\
      \hline
       $p-$value  & $5.31\times10^{-10}$ & $6.15\times10^{-17}$ & $1.89\times10^{-14}$\\
       \hline
    \end{tabular}
\end{table}

\section{Conclusions and Discussion}
\label{sec:conclusions}


In this study, we developed an individual-based model coupling internal cell cycle dynamics with population-level cell division and migration, to reproduce the key results from the laboratory experiments in Sect.~\ref{sec:experiment}, and to investigate the correlation between the cell cycle and wound closure. We investigated cell cycle arrest by minor model modifications and found that there is a significant impact on the emergent population dynamics when the arrest occurs at the G1 phase (i.e. $g$ state in the model), but not in the subsequent phases.

Additionally, we considered various initial conditions of the spatial pattern of cell states, which we observed had a significant impact on the healing process, specifically the closure rate. Hence, further investigation is required: from the experimental perspective, extracting such data will be relevant; from the modelling perspective, we can explore more specific patterns conditioned on the initial condition. This has the potential to suggest new experiments on treatments regarding the cell cycle phases to accelerate wound closure.

This brief study highlights the ability of the DLCM framework couple biologically motivated internal and population-level cell dynamics for comparison with \emph{in vitro} data. We also observe the more general utility of computational frameworks by using the model variants stated above to make mechanistic investigation that would, at best, demand far more resources in the biological lab.

Future research involves implementing various types of treatment such as electrical stimulation and light therapy, to reveal how they affect the cell cycle and subsequent wound closure rates, including an investigation of optimal does. Another potential direction is to add more detail to the DLCM model suggested by any discrepancy between simulation and data. Alternatively, we could utilize the framework model's close relation to a PDE system to model a larger number of cells or analyse the model's mean behaviour.

\begin{acknowledgement}
  EB and SE were supported by the Swedish Research Council under grant number 2019-03471 and by eSSENCE, a Swedish strategic research programme in e-Science. The work of QP was supported by Research England under the Expanding Excellence in England (E3) funding stream, which was awarded to MARS: Mathematics for AI in Real-world Systems in the School of Mathematical Sciences at Lancaster University. LP was supported by an EPSRC Industrial CASE Award with DSTL under grant number DSTLX1000152479R.
\end{acknowledgement}

\ethics{Competing Interests}{
The authors have no conflicts of interest to declare that are relevant to the content of this chapter.}

\ethics{Data Availability}{
The experimental data that support the findings of this study are available from RM and LP, upon reasonable request.
}


\bibliographystyle{abbrv}
\bibliography{main}

\end{document}